\documentclass[aps,prd,%showpacs,preprintnumbers,twelvepoint
 ]{revtex4}
\usepackage{graphicx}
\usepackage{color}
\usepackage{slashed}
\usepackage{amsmath}
\newcommand\bef{\begin{figure}}
\newcommand\eef[1]{\label{fg:#1}\end{figure}}
\newcommand\beq{\begin{equation}}
\newcommand\eeq[1]{\label{#1}\end{equation}}
\newcommand\beqa{\begin{eqnarray}}
\newcommand\eeqa[1]{\label{#1}\end{eqnarray}}
\newcommand\bet{\begin{table}}
\newcommand\eet[1]{\label{tb:#1}\end{table}}

\newcommand\fgn[1]{Figure \ref{fg:#1}}
\newcommand\eqn[1]{eq.\ (\ref{#1})}
\newcommand\scn[1]{Section \ref{sec:#1}}
\newcommand\apx[1]{Appendix \ref{sec:#1}}

\newcommand\tr{{\rm Tr\/}}

\newcommand\ie{{\sl i.e.\/}}

\newcommand{\I}{{\cal I}}

\newcommand{\N}{{\cal N}}
\newcommand{\bfq}{{\mathbf q}}

\newcommand{\bilin}[1]{\overline\psi{#1}\psi}
\newcommand{\V}[1]{{\bf #1}}

\newcommand\ppbar{\langle\overline\psi\psi\rangle}

\begin{document}

\title{An effective field theory for warm QCD}
\author{Sourendu\ \surname{Gupta}}
\email{sgupta@theory.tifr.res.in}
\affiliation{Department of Theoretical Physics, Tata Institute of Fundamental
         Research,\\ Homi Bhabha Road, Mumbai 400005, India.}
\author{Rishi\ \surname{Sharma}}
\email{rishi@theory.tifr.res.in}
\affiliation{Department of Theoretical Physics, Tata Institute of Fundamental
         Research,\\ Homi Bhabha Road, Mumbai 400005, India.}
\begin{abstract}
Using only global symmetries of QCD, we set up an effective model of
quarks at finite temperature near the cross over, including all possible
terms up to dimension 6.  We first treat this in mean field theory. Then
we investigate low-energy fluctuations around it up to one-loop order in
fermions below the cross over. Static correlation functions of pions and
the cross over temperature,  both measured on the lattice, completely
suffice to fix all parameters of the theory. We examine predictions of
this theory, including those for thermodynamic quantities.  The results
are encouraging.
\end{abstract}
\maketitle

\goodbreak\section{Introduction\label{sec:intro}}%\input{sec1.tex}

Effective field theories (EFTs) are good ways of organizing the
computation of low-energy or long-distance effects in a quantum field
theory. QCD at very low temperature, $T$, seems to be well described by
an EFT which describes the dynamics of pions \cite{wein,tchipt}. At very high
temperature it seems to be possible to understand long-distance phenomena
qualitatively in an effective weak-coupling expansion \cite{kari}. These
rely on a separation of scales $T\gg gT\gg g^2T$, where $g$ is the gauge
coupling at momentum scale of order $T$. The EFT at each different momentum
scale is obtained by integrating over momenta larger than those required
at this scale. The separation of scales breaks down when $T$ is a few
hundred MeV, since $g\sim1$ in that temperature range, and these EFTs
also become ineffective.

However, this is precisely the range of temperatures which is of great
physical interest. The finite temperature cross-over from a chiral
symmetry broken hadronic state to a symmetry restored quark-gluon state
occurs here. It is also the range of temperatures which seems to be most
relevant for experiments using heavy-ion collisions. Some models have
been used to explore the physics of this region and have had moderate
success in matching lattice computations \cite{njl,pnjl}.

Here we investigate a related model EFT which is built to describe the
physics of QCD at finite temperature, around the cross over temperature
$T_{co}$.  We follow a method which is often used when there is very
limited information about the system under study \cite{wein}. Then one can
write an EFT by guessing what the relevant fields and global symmetries
are, and then writing down all possible terms in the Lagrangian which
are constrained by the relevant global symmetries. Since we already know
the full theory, namely QCD, it may seem that this process is less than
optimal, since it does not exploit all the knowledge that the theory
contains. The problem in deriving an effective theory from QCD, however,
is that there is no known small parameter which can be exploited to do
this accurately.

The model is written down in \scn{eft}. The mean field theory (MFT) is
briefly examined in \scn{mft}. Fluctuations around the MFT are considered
in \scn{fluct}. The description of lattice measurements is taken up in
\scn{fit}. The regularization of integrals is discussed in \apx{reg}.
These are technical parts of the paper. A non-technical discussion is
in the concluding \scn{disc}; it is possible to read this section before
the rest of the paper.

\goodbreak\section{The effective theory model\label{sec:eft}}%\input{sec2.tex}

The global symmetries of QCD which we use to build an EFT model are
the vector (V) and axial (A) flavour symmetries SU$_V$($N_f$) $\times$
SU$_A$($N_f$), for $N_f$ flavours of quarks.  We will build a model
of interacting quark fields designed to match physics near some finite
temperature $T_0$. The model is organized by mass dimension of operators,
using the intrinsic scale $T_0$ to give dimensions to couplings where
necessary.  The fermion field carry Dirac and flavour indices. We carry
along the SU($N_c$) colour index, although these contribute only overall
factors since there are no colour interactions in the model: every fermion
bilinear we build is colour blind.  We have no a priori argument for this,
but proceed on this assumption to examine phenomena. We use the notation
$\N=4N_cN_f$ for the dimension of the fermion field, choosing $N_c=3$
and $N_f=2$ in this paper. The extension to $N_f=3$ is interesting and 
will be examined in future.

Euclidean Dirac matrices are chosen to be Hermitean, with $\gamma_4 = -i
\gamma_0$, and $\gamma_5=\gamma_1\gamma_2\gamma_3\gamma_4$.  The conjugate
Fermion field is defined as $\overline\psi=\psi^\dagger\gamma_4$.
Since the Lorentz group becomes a rotation group, its generators are
Hermitean, $S_{\mu\nu}=-i[\gamma_\mu,\gamma_\nu]/4$. Since we model
thermal physics, time and space components are distingushed. As a
result the finite temperature theory breaks the full O(4) rotational
symmetry to a cylidrical symmetry O(3)$\times$Z$_2$, where O(3) is the
spatial rotational symmetry and Z$_2$ is the Euclidean time reversal
symmetry T \cite{symm}. Every O(4) tensor also reduces.  A finite
temperature effective field theory then has many more couplings than a
zero temperature theory. In order to describe the thermodynamics of the
original theory, it suffices to write only CPT invariant terms in the EFT.

We will write the Lagrangian of the EFT without a vacuum energy term,
$d^0T_0^4$. There are no terms of dimension one or two. The dimension
three operator $\bilin{}$ has the quark pole mass as its coefficient,
which we write as $m_0=d^3T^0$. The dimension four terms are obtained
by using derivative operators: $\bilin{\slashed\partial_0}$ and
$\bilin{\slashed\partial}$.  Here $\slashed\partial_0=\gamma_0\partial_0$
and $\slashed\partial=\gamma_i\partial_i$.  These are invariant under P
and T, and also V and A. Inserting further Dirac structures in the two
bilinears makes them lose discrete symmetries. The coefficient of the
kinetic term, $\bilin{\slashed\partial_0}$ fixes the normalization of
the field operator, and hence is always set to unity. The coefficient
of the other term, $d^4$, is special to finite temperature.

In a quadratic theory of fermions, the poles of the inverse propagator
would be the solution of $p_0^2+(d^4)^2|{\bf p}|^2+m_0^2=0$.  Pole mass,
$im_0$, is the term used for the pole of the temporal correlator
averaged over space, \ie, for $|{\bf p}|=0$. The screening mass is
the pole of the static propagator, \ie, for $p_0=0$.
So $d^4\ne1$ is just the statement that
screening and pole masses are not equal. A general limit on $d^4$ comes
from the requirement that after Wick rotation the group velocity of a wave packet should
be less than unity. In a quadratic theory this implies that $0<d^4<1$.
With an UV cutoff, one can have larger values of $d^4$ without running
into problems with causality.

Three terms of mass dimension five can be obtained by using
derivatives: $\bilin{\slashed\partial_0\slashed\partial_0}$,
$\bilin{\slashed\partial_0\slashed\partial}$, and
$\bilin{\slashed\partial\slashed\partial}$.  All three are invariant under
C, P, T and V, but not under A. One can restore A by putting extra Dirac
matrices in the bilinears, but this will destroy some of the discrete
spacetime symmetries. As a result, there are no dimension five terms in
the model.

Terms of mass dimension six can either be built
using fermionic current-current interactions or using
derivatives in fermion bilinears. Four terms of the
second kind are $\bilin{\slashed\partial_0^3}$,
$\bilin{\slashed\partial_0^2\slashed\partial}$,
$\bilin{\slashed\partial_0\slashed\partial^2}$, and
$\bilin{\slashed\partial^3}$.  All three are invariant under P and T,
V and A. One can use the equation of motion at dimension 4 to remove
$\slashed\partial_0$ and reduce all terms to that involving only spatial
derivatives. This term should be added to the effective theory. However,
it affects neither the MFT nor the fluctuations at the order we examine.
So we do not include it here. Inserting further Dirac structures in
the two bilinears makes them lose the invariance under P, C or T. The
four-Fermi terms are restricted by chiral invariance. We may add terms
of higher dimension if required.

The Euclidean EFT model we start with then consists of all possible terms
up to mass dimension six, invariant under the global and space-time symmetries
of a finite temperature Euclidean theory,
\beqa
\nonumber
  L &=& 
        d^3T_0\bilin{}
      + \bilin{\slashed\partial_4} %- \mu \bilin{\gamma_4}
      + d^4 \bilin{\slashed\partial_i} 
      + L_6 \qquad{\rm where}\\
\nonumber 
  L_6 &=&
      + \frac{d^{61}}{T_0^2} \left[(\bilin{})^2 
            + (\bilin{i\gamma_5\tau^a})^2 \right]
      + \frac{d^{62}}{T_0^2} \left[(\bilin{\tau^a})^2 
            + (\bilin{i\gamma_5})^2 \right] \\
\nonumber &&
      + \frac{d^{63}}{T_0^2} (\bilin{\gamma_4})^2 
      + \frac{d^{64}}{T_0^2} (\bilin{i\gamma_i})^2 
      + \frac{d^{65}}{T_0^2} (\bilin{\gamma_5\gamma_4})^2 
      + \frac{d^{66}}{T_0^2} (\bilin{i\gamma_5\gamma_i})^2  \\
\nonumber &&
      + \frac{d^{67}}{T_0^2} \left[(\bilin{\gamma_4\tau^a})^2 
            + (\bilin{\gamma_5\gamma_4\tau^a})^2 \right]
      + \frac{d^{68}}{T_0^2} \left[(\bilin{i\gamma_i\tau^a})^2 
            + (\bilin{i\gamma_5\gamma_i\tau^a})^2\right]  \\
&&    + \frac{d^{69}}{T_0^2} \left[(\bilin{iS_{i4}})^2 
            + (\bilin{S_{ij}\tau^a})^2 \right]
      + \frac{d^{60}}{T_0^2} \left[(\bilin{iS_{i4}\tau^a})^2 
            + (\bilin{S_{ij}})^2 \right]
\eeqa{eftaction}
This differs from the NJL model \cite{njl} in a few ways. First, it is
built to model QCD at finite temperature, hence Lorentz invariance is not
used, and a temperature scale $T_0$ is used to organize the expansion.
Second, it is an EFT, so all terms up to a certain order in mass dimension
are kept, provided they are invariant under the symmetries of the model.
The NJL model would have all four-fermi couplings set to zero except
$d^{61}$. Some of the other couplings have been considered before \cite{terms}.

For later use we point out a fact which is well-known \cite{weinberg}. Since the
dimension-6 terms are specifically written to preserve both the vector
and axial parts of the chiral symmetry, it is only the mass term which
breaks the symmetry. We have chosen the mass matrix to be diagonal in
flavour, so only the axial part of the chiral symmetry is broken by
it. As a result, we have the following relation for the divergence of the
axial current,
\beq
   \partial_\mu j^a_\mu(x)=2 d^3T_0 P^a, \qquad{\rm where}\qquad
    P^a = \frac12\overline\psi\gamma_5\tau^a\psi.
\eeq{pionop}
This is the partially conserved axial current (PCAC) relation.

\goodbreak\section{The mean field theory\label{sec:mft}}%\input{sec3.tex}

The fermionic mean-field approximation is the operator identity
  $\overline\psi_\alpha\psi_\beta = \delta_{\alpha\beta}\ppbar$,
where $\alpha$ and $\beta$ are composite Dirac-flavour-colour indices.
Performing the Wick-contractions in various ways in the generic 4-fermi term
then gives
\beq
 (\bilin{\Gamma})^2=2\ppbar\left[\tr\Gamma\bilin{\Gamma}
   -\bilin{\Gamma\Gamma}\right]
   -\ppbar^2\left[(\tr\Gamma)^2-\tr(\Gamma\Gamma)\right].
\eeq{fierz}
The product of Dirac-flavour matrices in the second term is the Fierz
transformation. Since all the Dirac matrices as well as the rotation
generators are traceless, only $\Gamma=1$ contributes to the first term.
Also, since $\gamma_\mu^2=1$ for all Euclidean Dirac matrices, we find
that the second is proportional to the identity for all currents. 
Using this we can reduce the interactions to an effective mass term
\beqa
\nonumber
L_6 &=& -\N\left(\frac{T_0^2}{4\lambda}\right)\Sigma^2
    + \Sigma\bilin{}, \qquad{\rm where}\qquad
   \Sigma=\frac{2\lambda}{T_0^2}\,\ppbar\qquad{\rm and}\\
 &&\quad
   \lambda=(\N+2)d^{61}-2d^{62}-d^{63}+d^{64}+d^{65}-d^{66}
      +d^{69}-d^{60}.
\eeqa{dim6}
Note that exactly the same result would have been obtained in the mean
field approximation to the NJL model \cite{njl}. The NJL mean field
theory is widely explored \cite{klevansky}, and we only need to adapt the
results to take into account Lorentz symmetry breaking. The EFT is
\beq
  L_{{\scriptscriptstyle{\rm MFT}}} =
       -\N\left(\frac{T_0^2}{4\lambda}\right)\Sigma^2
      + \bilin{\slashed\partial_4} %- \mu \bilin{\gamma_4}
      + d^4 \bilin{\slashed\partial_i} + m\bilin{}
    \quad{\rm where}\quad m=d^3T_0+\Sigma
\eeq{emft} is an effective quark mass. $T_0$ can be chosen as we wish.
There are only three couplings in this model. These have to be determined
from data. $\Sigma$ must come out of a computation. We use dimensional
regularization to deal with this theory. Details are given in
\apx{reg}, which also contains the notation used in the remainder of
the paper.

\subsection{$T_0$ and couplings}

Using the methods of \apx{reg}, we find that the free energy density in
the MFT is
\beq
   -\Omega = \frac{\N T_0^2\Sigma^2}{4\lambda} 
     + \frac{\N m^4}{64\pi^2(d^4)^3}\left[\log\left(
          \frac{m^2}{(d^4)^2M^2}\right)-\frac32\right]
     + \frac{\N T}{2\pi^2(d^4)^3}\int_0^\infty dp p^2\log\left[
                  1+\exp\left(-\frac ET\right)\right],
\eeq{pmft}
where $M$ is a scale from dimensional regularization and $E^2=p^2+m^2$,
where $p$ is rescaled in the last term to include $d^4$, so giving rise
to the Jacobian factor $1/(d^4)^3$.  The value of $\Sigma$ must be that
which minimizes $\Omega$ for fixed $T$. This, and the gap equation,
$d\Omega/d\Sigma=0$, are very similar to the usual solutions of the NJL
model. Putting the value of $\Sigma$ at the minimum back into \eqn{pmft}
one obtains the free energy density as a function of temperature,
$\Omega_0$. The pressure is $P=-\Omega_0$.

In the chiral limit and at low temperature, the minimum is at
non-vanishing $\Sigma$, whereas at high temperature $\Sigma=0$ is the
only solution. The temperature at which the trivial solution becomes
the minimum is the critical temperature, $T_c$. By taking the second
derivative of $\Omega$ with respect to $\Sigma$ and requesting that this
vanish at $T_c$, one finds

\beq
   \frac{(d^4)^3}\lambda = \frac1{12\pi^2}\; \frac{T_c^2}{T_0^2}.
\eeq{deftc}
Since we are interested in the region where the system crosses over from
one state to another, we can choose $T_0$ to be equal to $T_c$. In this
case $\lambda = 12\pi^2(d^4)^3$. Note that this is just a convention for
$T_0$, and not a prediction of $T_c$. By changing the convention we would
only shift the value of $\lambda/ (d^4)^3$, while keeping $T_c$ unchanged.

Note that there are two more couplings to be determined, namely $d^3$ and
$d^4$. We will have to use two observables in order to fix these.  Note
that unlike the combination $\lambda/ (d^4)^3$, these may depend on the
renormalization scale $M$. On changing $M$ one needs to change couplings
in order to keep the observables unchanged. This is a renormalization
group running for the couplings, although its validity is limited because
$M$ cannot be made arbitrarily large. It may seem that a freedom to choose
the coupling $\lambda$ to fit a third observation has gone away. This
is not so; the freedom has been transmuted into a choice of the as yet
unspecified dimensionful quantity $T_0$. A third observation is required
to fix this coupling. Once this is done, everything else is a prediction.

\subsection{Curvature of the critical line}

One may add a chemical potential term to the action in the form
$-\mu\bilin{\gamma_4}$. It turns out then that the critical point in
the chiral limit at $\mu=0$ develops into a critical line. The same
computation as above, now done at small $\mu$ gives an equation for the
critical line,
\beq
  T_c(\mu)^2 +\frac 3{\pi^2}\;\mu^2=T_0^2,
\eeq{criticalline}
in the chosen convention $T_0=T_c(0)$ and quite independent of the
couplings in the theory.  The curvature of the critical line in the
chiral limit is usually given in terms of the expansion
\beq
   T_c(\mu) = T_c(0) - \frac12\kappa\;\frac{\mu^2}{T_c(0)}
      +{\cal O}(\mu^3).
\eeq{defkappa}
Comparing these two equations, we find the parameter free prediction
\beq
  T_c(0)\kappa=\frac3{\pi^2}.
\eeq{curvcrit}
Estimates of this quantity have been made on the lattice with quarks which
are somewhat heavier than those found in nature. The results correspond
to the range $T_c(0)\kappa\simeq0.01$--0.05 \cite{kappa}. Although the
mean field prediction in the chiral limit is larger, it is not so far
away that it cannot be improved by various corrections.

These include corrections due to  fluctuations. Interestingly though,
there are other corrections which may be larger. The perturbation to
the action due to a chemical potential breaks CP symmetry by a dimension
3 term. This hard breaking in the UV could generate other CP-violating
terms in the effective action. Enumerating and controlling them all is
a problem we will address in the future.

\goodbreak\section{Pionic fluctuations\label{sec:fluct}}%\input{sec4.tex}

\subsection{The pion field}

In order to examine fluctuations around the mean field solution of the
fermionic model, we use the Hubbard-Stratanovich trick and introduce a
matrix valued field $V$ with composite Dirac, flavour and colour indices
to linearize the action in \eqn{eftaction}. The equation of motion for
$V$ gives
\beq
   V_{\alpha\beta} = q_\alpha\overline q_\beta.
\eeq{hst}
Since the dimension-6 terms of the action have been constructed
to be invariant under axial flavour transformations, simultaneous
transformations of $V$, $\psi$ and $\overline\psi$ leave even the
linearized form invariant. 

Fluctuations in the axial direction about the condensate are therefore
captured by local ``isospin-waves'' parametrized in the form
\beq
   \psi\to{\rm e}^{i\pi^a\tau^a\gamma_5/(2f)}\psi \qquad{\rm and}\qquad
   \overline\psi\to\overline\psi{\rm e}^{i\pi^a\tau^a\gamma_5/(2f)},
\eeq{transf}
where $\pi^a$ are bosonic fields, and $f$ an emergent constant of
dimension one.  These fields drop out of the dimension-6 terms, and are
seen only in the dimension-4 terms. Since the path-integral is now
quadratic in the fermion fields, we integrate them out to one-loop
order to get the tree-level action of the fluctuations up to dimension-4
\beq
  L_f = \frac{c^2 T_0^2}2\pi^2 + \frac12(\partial_0\pi)^2 
    + \frac{c^4}2(\nabla\pi)^2 + \frac{c^{41}}8\pi^4.
\eeq{quadpi}
The only possible contraction of SU(2) flavour indices for the quartic
term is $(\pi^2)^2$.  A similar theory without the quartic terms has been
previously considered in \cite{ss}; in their notation $c^2T_0^2=m^2$
and $c^4=u^2$. The theory seems to contain four constants, namely $f$,
and the couplings $c^2$, $c^4$ and $c^{41}$. However, the underlying
fermionic theory from which we derive this has only three couplings.
In a later section we will give the computation
of the couplings in \eqn{quadpi} to one-loop order.

The contribution of the fluctuations to the free energy from the
tree-level action of \eqn{quadpi} is straightforward, since it is
reduces to a computation only in the quadratic theory. Since this closely
parallels the MFT computation, we write down the result
\beq
   \Omega_\pi = \frac{3(c^2T_0^2)^2}{64\pi^2(c^4)^{3/2}}\left[
      \log\left(\frac{c^2T_0^2}{c^4M^2}\right)-\frac32\right]
      + 3T\int\frac{d^3p}{(2\pi)^3}\,\log\left(1-{\rm e}^{-E/T}\right)
\eeq{pifene}
where $E^2=c^4p^2+c^2T_0^2$ and $M$ is the scale which arises from
dimensional regularization of the vacuum energy term. The thermal
integral is well-known.

\subsection{Current algebra and parameters of the quadratic action}

We first make a few remarks about symmetry relations using \eqn{quadpi}.
At small momentum we have
\beq
\int d^4x {\rm e}^{iq\cdot x}\langle\pi^a(x)\pi^b(0)\rangle 
   = \frac{\delta^{ab}}{c^2T_0^2 + q_4^2+ c^4\bfq^2}
\eeq{pioncor}
In the static limit one obtains
\beq
  \lim_{q^4\rightarrow 0}\int d^4x{\rm e}^{iq\cdot x}
   \langle\pi^a(x)\pi^b(0)\rangle 
   = \frac{\delta^{ab}}{c^4}\times\frac1{\bfq^2+(c^2T_0^2/c^4)}
\eeq{pipi}
This implies that the pion screening mass is $M_\pi=T_0\sqrt{c^2/c^4}$. Use
of an effective theory simplifies the computation of screening masses, a
fact that has been used before \cite{scrm}.

Using the definition of the axial vector rotation by an angle $\theta^a$ on
quark fields, and the definition of the pion field, one sees
that the transformation acts on the pion field as
\beq
  \pi^a(x) \longrightarrow \pi^a(x) + 2f\theta^a
\eeq{tranfpi}
The axial Noether currents are then
\beq
  j_4^a(x) = 2f\partial_4\pi^a + \cdots\qquad
  j_i^a(x) = 2f\,c^4\partial_i\pi^a +\cdots.
\eeq{noether}
where the dots represent terms that arise when higher dimensional terms are
kept in the action. As a result, current correlators can be written as
\beqa
\nonumber
\langle j_4^a(x)j_4^b(y)\rangle &=& (2f)^2
   \langle\partial_4\pi^a(x)\partial_4\pi^b(y)\rangle + \cdots\\
\langle j_i^a(x)j_i^b(y)\rangle &=& (2f\,c^4)^2
   \langle\partial_i\pi^a(x)\partial_i\pi^b(y)\rangle + \cdots
\eeqa{corrs}

The momentum space correlator of the axial charge density can now be
written as
\beq
\int d^4x{\rm e}^{iq\cdot x}\langle j_4^a(x)j_4^b(0)\rangle =
  (2f)^2\delta^{ab}q_4^2 \langle\pi^a(q_4,\bfq)\pi^b(-q_4,-\bfq)\rangle
  = \frac{(2f)^2\delta^{ab}q_4^2}{c^2T_0^2+q_4^2+c^4\bfq^2}
   \stackrel{q_4\to0}\longrightarrow0.
\eeq{qcorrel}
Since the static limit gives a vanishing result, screening correlators of
this component of the current cannot be used to constrain the parameters
of the effective theory.

For the correlator of the spatial part of the current, we have
\beq
\int d^4x{\rm e}^{iq\cdot x}\langle j_i^a(x)j_i^b(0)\rangle =
  (2f c^4)^2\delta^{ab}\bfq^2 \langle\pi^a(q_4,\bfq)\pi^b(-q_4,-\bfq)\rangle
  = \frac{(2f c^4)^2\delta^{ab}\bfq^2}{c^2T_0^2+q_4^2+c^4\bfq^2}
   \stackrel{q_4\to0}\longrightarrow
     \frac{(2f)^2c^4\delta^{ab}\bfq^2}{\bfq^2+M_\pi^2}
\eeq{jcorrel}
This differs from the pion correlator only in the coefficient. Since the
measurement of a single current correlation function can give both the
coefficient and the screening mass, we can constrain the model entirely
by the measurement of this correlation function.

Given the relations in \eqn{pionop} and \eqn{noether}, one easily finds
\beq
  \int d^4x{\rm e}^{iq\cdot x} \langle P^a(x)P^b(0)\rangle 
   = \left(\frac f{2m_0}\right)^2\,\delta^{ab}
     \frac{(q_4^2+c^4\bfq^2)^2}{c^2T_0^2+q_4^2+c^4\bfq^2}
   \stackrel{q_4\to0}\longrightarrow
     \frac{f^2c^4}{4m_0^2}\times\frac{\bfq^4\delta^{ab}}{\bfq^2+M_\pi^2}.
\eeq{pp}
So the measurement of the static correlator for $P^a$ can also be
used to find both the screening mass and a combination of the
couplings.

It is useful to convert this to a form which can be directly implemented
on the lattice. As long as the lattice cutoff is sufficiently larger than
$T_0$, one can ignore the lattice spacing and treat it as a theory in a
finite box. The screening correlators are usually projected on to zero
momentum in the transverse direction, and measured in a cubic lattice
of side $L$. Then the screening correlator for $P^a$ is easily found to be
\beq
  C_P(z) = \frac{f^2c^4M_\pi^3}{4m_0^2}\,{\rm e}^{-M_\pi L/2}
     \cosh\left[M_\pi\left(\frac L2-z\right)\right].
\eeq{ppscr}
With the aid of a finite temperature version of the Gell-Mann-Oakes-Renner
relation, which we show later, this can be put into a form which is
directly measurable on the lattice.
The screening correlator for the axial current polarized in the $z$-direction
takes on the simple form,
\beq
  C_{J_3}(z) = \frac12 f^2c^4M_\pi\,{\rm e}^{-M_\pi L/2}
     \cosh\left[M_\pi\left(\frac L2-z\right)\right].
\eeq{jjscr}
This is also measurable on the lattice.

\subsection{Two-point functions}

\bef
\includegraphics[scale=0.25]{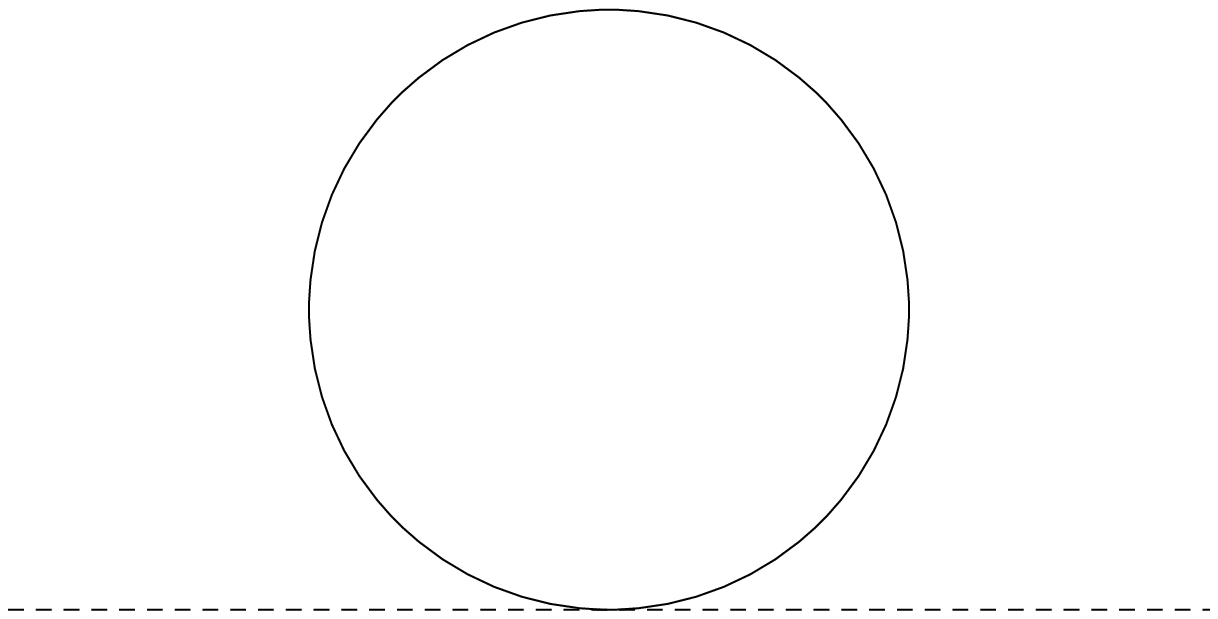}
\hspace{1cm}
\includegraphics[scale=0.25]{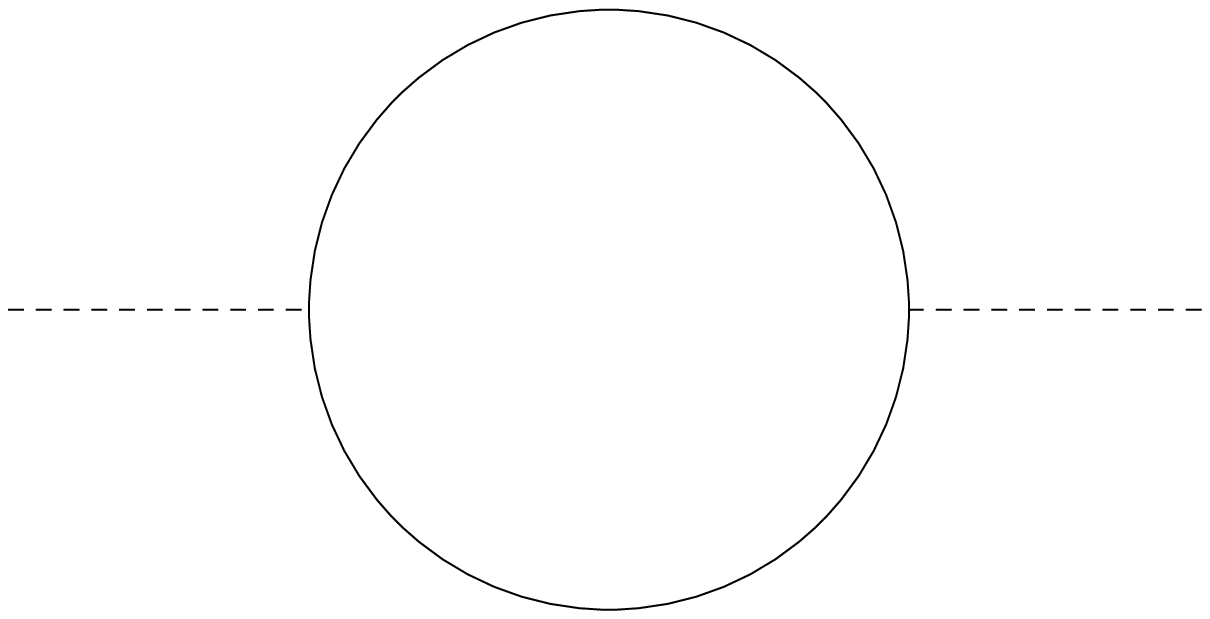}
\caption{Feynman diagram topologies for the two point function of pions
 obtained by integrating out the quarks. Dashed lines stand for pions,
 full lines for quarks.}
\eef{2pt}

We introduce the trasformations \eqn{transf} into the MFT action, and
then expand out the exponential to the lowest order needed for each of
the following computations. The couplings needed in \eqn{quadpi} can be
obtained by computing the two-point and four-point functions using the
processes shown in \fgn{2pt} and \fgn{4pt} respectively. These Feynman
diagrams can be evaluated using the quark propagator obtained from the
MFT and pion-quark couplings from the expansion of the exponential as
explained here.

The two-point function can be written in momentum space as,
\beq
 S_\pi = \frac{\N}{8f^2} \int\frac{d^4q}{(2\pi)^4}\;
  \widetilde\pi^a(q_\mu)\widetilde\pi^a(-q_\mu)
  [-q_4q_4\I_{44}(q_\mu) -q_iq_i\I_{ii}(q_\mu)+\I(q_\mu)],
\eeq{Spi} 
where the integrals come from the one-loop Feynman diagrams shown
in \fgn{2pt}. The explicit form of the integrals are
\beqa
\nonumber
 \I_{44}(q)&=&\frac14\int\frac{d^4p}{(2\pi)^4}\;\tr\left[
    \frac1{-i\slashed p+m} (\gamma^5\gamma^4)
    \frac1{-i\slashed p'+m} (\gamma^5\gamma^4)\right],\\
\nonumber
 \I_{ii}(q)&=&\frac{(d^4)^2}4\int\frac{d^4p}{(2\pi)^4}\;\tr\left[
    \frac1{-i\slashed p+m} (\gamma^5\gamma^i)
    \frac1{-i\slashed p'+m} (\gamma^5\gamma^i)\right],\\
\nonumber
 \I(q)&=&\frac{m_0^2}4\int\frac{d^4p}{(2\pi)^4}\;\tr\left[
    \frac1{-i\slashed p+m} (i\gamma^5)
    \frac1{-i\slashed p'+m} (i\gamma^5)\right]\\
  &&+\frac{m_0}4\int\frac{d^4p}{(2\pi)^4}\;\tr\left[
    \frac1{-i\slashed p+m}\right].
\eeqa{oneloopints}
where $p'=p+q$ and the trace is over spinor indices.

Taking the limit $q_\mu\to0$ in the integrals allows us to match
$S_\pi$ to the effective action in \eqn{quadpi}.
This identification
then implies the matching conditions
\beq
   f^2=-\frac{\N}4I_{44}(0),\quad
   c^4=\frac{\I_{ii}(0)}{\I_{44}(0)}, \quad
   c^2T_0^2=-\,\frac{4\I(0)}{\I_{44}(0)}.
\eeq{matching}

We write $p=(p^4,{\bf p})$ and
recall that $p^4=(2n+1)\pi T$. Using the definition of the sum-integral,
as before, we have
\beqa
\nonumber
 \I_{44}(0)&=&-T\sum_{n=-\infty}^\infty\int\frac{d^3p}{(2\pi)^3}\;
     \frac{(p^4)^2-(d^4)^2\V p^2+m^2}{(p^2+m^2)^2}\\
\nonumber
 \I_{ii}(0)&=&-(d^4)^2T\sum_{n=-\infty}^\infty\int\frac{d^3p}{(2\pi)^3}\;
     \frac{2(d^4)^2\V p_i\V p_i-(p^4)^2-(d^4)^2\V p^2+m^2}
      {(p^2+m^2)^2}\\
 \I(0)&=&-m_0(m_0-m)T\sum_{n=-\infty}^\infty\int\frac{d^3p}{(2\pi)^3}\;
     \frac1{p^2+m^2}
\eeqa{Iq0formal}

After performing the sums in $p^4$ using the methods of \cite{kapusta} the integrands can be split into a 
thermal part and a part that is independent of $T$. The thermal parts
of the integrals are exponentially damped for momenta much larger than
$T$, and hence have no UV divergences. However, they cannot be exhibited
in closed form, and are best evaluated numerically. We find that 
the integrals are
\beqa
\nonumber
  \I_{44} &=&  \frac{m^2}{4\pi^2(d^4)^3}\;\log\left(\frac m{d^4M}\right)
       +\int\frac{d^3p}{(2\pi)^3} \left[\frac{m^2}{E^3(\exp(E/T)+1)}
    -\frac{(d^4)^2\V p^2\exp(E/T)}{E^2T(\exp(E/T)+1)^2}\right]\\
  \\
\nonumber
  \I_{ii} &=&  \frac{m^2}{8\pi^2d^4}\left[4\log\left(\frac m{d^4M}\right)-1\right]
       -\int\frac{d^3p}{(2\pi)^3}
     \left[\frac{(d^4)^4\V p^2(1-\cos^2\theta)}{E^3(\exp(E/T)+1)}
    -\frac{((d^4)^4\V p^2\cos^2\theta+(d^4)^2m^2)\exp(E/T)}
      {E^2T(\exp(E/T)+1)^2}\right]\\
  \I &=&  \frac{m_0(m_0-m)m^2}{16\pi^2(d^4)^3}\left[1-2\log\left(\frac m{d^4M}\right)\right] +
       \int\frac{d^3p}{(2\pi)^3}
     \frac{m_0(m_0-m)}{E(\exp(E/2T)+1)}.
\eeqa{I44IiiI}
The vacuum terms are regulated in DR and $M$ is the $\overline{MS}$
scale defined in \apx{reg}.

It can be shown that $c^4$ vanishes at $T_c$ in the chiral limit. Since the
vacuum part of $\I_{ii}$ vanishes for $m\rightarrow0$, we obtain
\beq
 \I_{ii}(0) = -\frac1{6\pi^2d^4}
   \int_0^{\infty}\frac{dp\,p^2}{\exp(p/T)+1}\left[\frac2p-
     \frac1{T\{\exp(-p/T)+1\}}\right].
\eeq{IDimReglimw0q0mu0m0}
In terms of the variable $y=p/T$, the integral can be written as
\beq
 \I_{ii}(0) = -\frac{T^2}{6\pi^2d^4}(I_1-I_2),
      \quad{\rm where}\quad
   I_1=2\int_0^\infty\frac{dy\,y}{{\rm e}^y+1},\quad
   I_2=\int_0^\infty\frac{dy\,y^2}{(1+{\rm e}^y)(1+{\rm e}^{-y})}.
\eeq{finalcpi}
Interestingly, both $I_1$ and $I_2$ are equal to $\pi^2/6$, as a result
of which $\I_{ii}(0)$ vanishes. The cancellation can also be seen in the
following way. $I_1$ has a set of single poles, and $I_2$ has double
poles at exactly the same set of points. Furthermore, the residues at
the poles exactly cancel between the two integrals. Since $\I_{ii}(0)=0$,
by \eqn{matching} $c_4$ vanishes at $T_c$ in the chiral limit \cite{ss}.
Then, from the equation of motion, one sees that the pion field just
becomes a constant non-propagating field: the pion disappears above the
critical point in the chiral limit. This creates no pathologies for the
1-loop computation. However, in pion loops, thermal integrals are not
UV regulated. This means that near $T_c$ higher dimensional terms must
be added to \eqn{quadpi} in order to make sense of the pion EFT.

Within the 1-loop computation it is possible to ask how $c^4$ approaches
zero near the chiral critical point. The argument above implies that
the thermal part of $\I_{ii}$ varies as $m^2=\Sigma^2$. However, when
$\Sigma$ is not exactly zero, the $T=0$ part must also be taken into
account, and adds a $m^2\log m$ term. At the same time, $\I_{44}$ goes
to a constant.  From the gap equation we see that $\Sigma$ goes to zero
at $T_c$ with a power behaviour $\sqrt{1-T/T_c}$. However, a more careful
examination shows that this is modified by a $\sqrt{1-T/T_c}\log(1-T/T_c)$
term. If one drops the logarithms, one would see $c^4\propto(1-T/T_c)$
\cite{ss}. The logarithms make this vanish slower, giving a leading
behaviour $c^4\propto (1-T/T_c)\log(1-T/T_c)$.

\subsection{Thermal Gell-Mann-Oakes-Renner relation}

An useful formal relation is called the Gell-Mann-Oakes-Renner (GMOR)
relation.  For small $m_0$ in \eqn{Iq0formal}, we can neglect the term
quadratic in $m_0$, and write
\beq
 \I(0) = 2m_0\int\frac{d^4p}{(2\pi)^4}\;\frac m{p^2+m^2}
\eeq{gortry}
where the overall factor of 2 on the right comes from the fact that the
4-d integral is defined by a sum over positive Matsubara modes, but
$\I$ is defined by a sum over all Matsubara modes.
Next, using \eqn{matching} we obtain,
\begin{equation}
 c^2T_0^2 =  \frac{\N}{f^2} \I(0)
   =\frac{2m_0\N}{f^2}\int\frac{d^4p}{(2\pi)^4}\;\frac m{p^2+m^2}
\end{equation}
The gap equation in the MFT shows that the integral is exactly $\ppbar/2$.
As a result, we obtain the extension of the GMOR relation to finite
temperature,
\beq
 c^2T_0^2 = -\,\frac{\N m_0\ppbar}{f^2}.
\eeq{gor}
The factor of $\N$ is conventional; it is often absorbed into the
definition of the condensate.  Since all the quantities appearing here
have been defined in terms of the couplings in the fermion theory, this
identity is a statement of self-consistency at one-loop order. Note that
when $m_0=0$ the identity implies $c^2=0$.

\subsection{Four-point functions}

\bef
\includegraphics[scale=0.25]{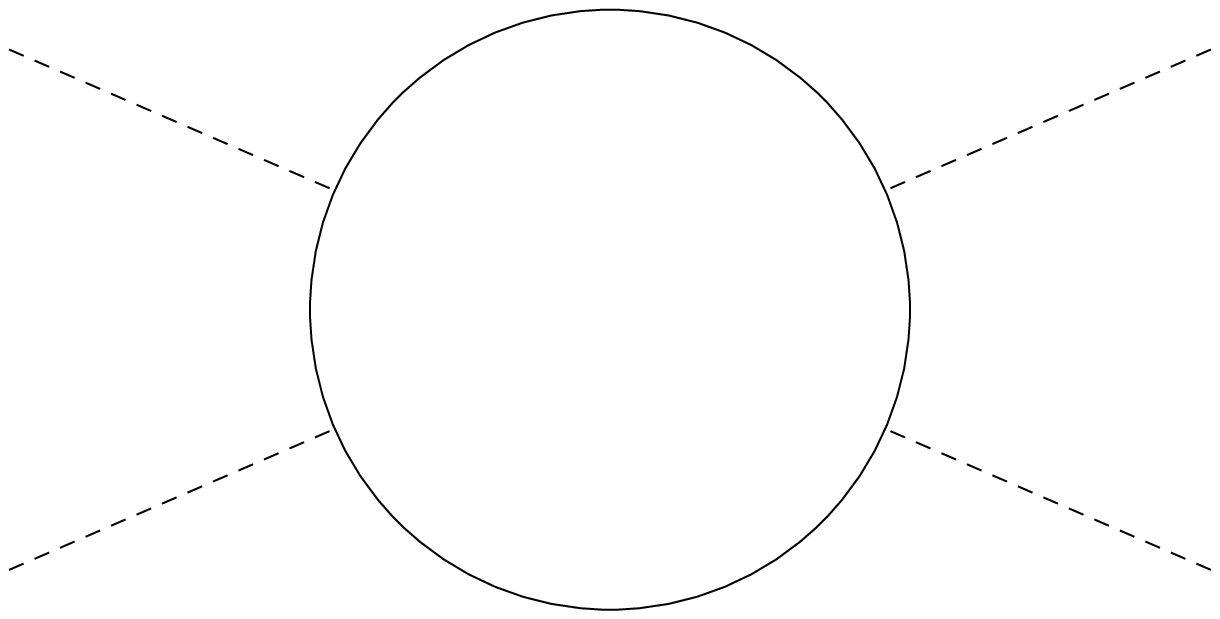}
\hspace{2mm}
\includegraphics[scale=0.25]{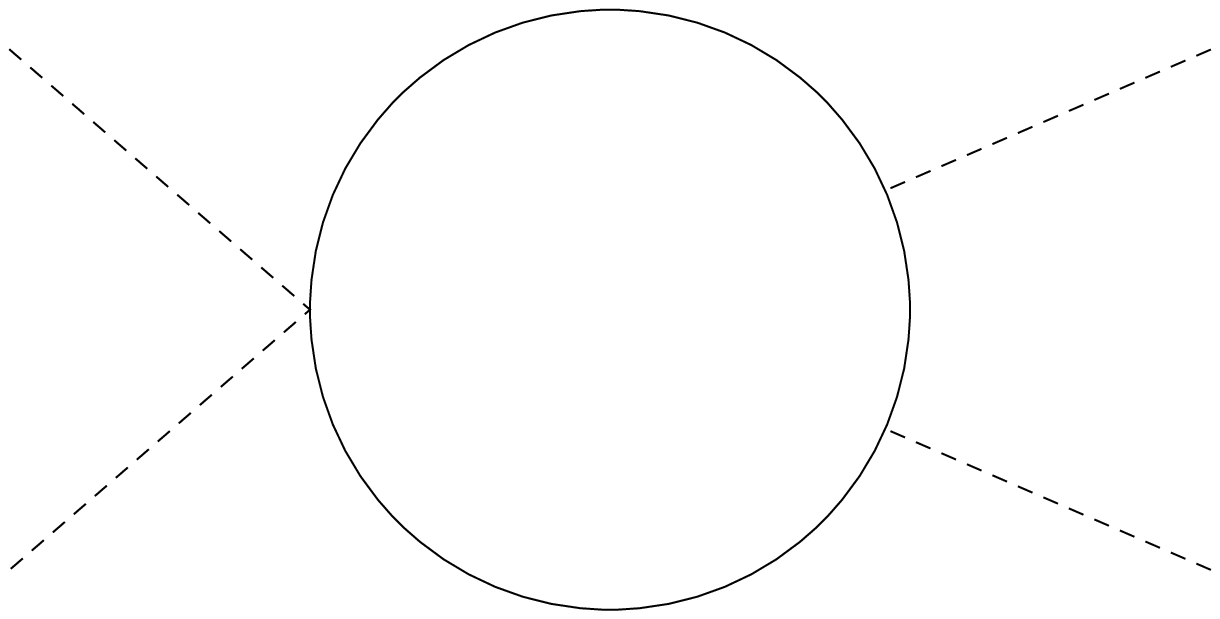}
\hspace{2mm}
\includegraphics[scale=0.25]{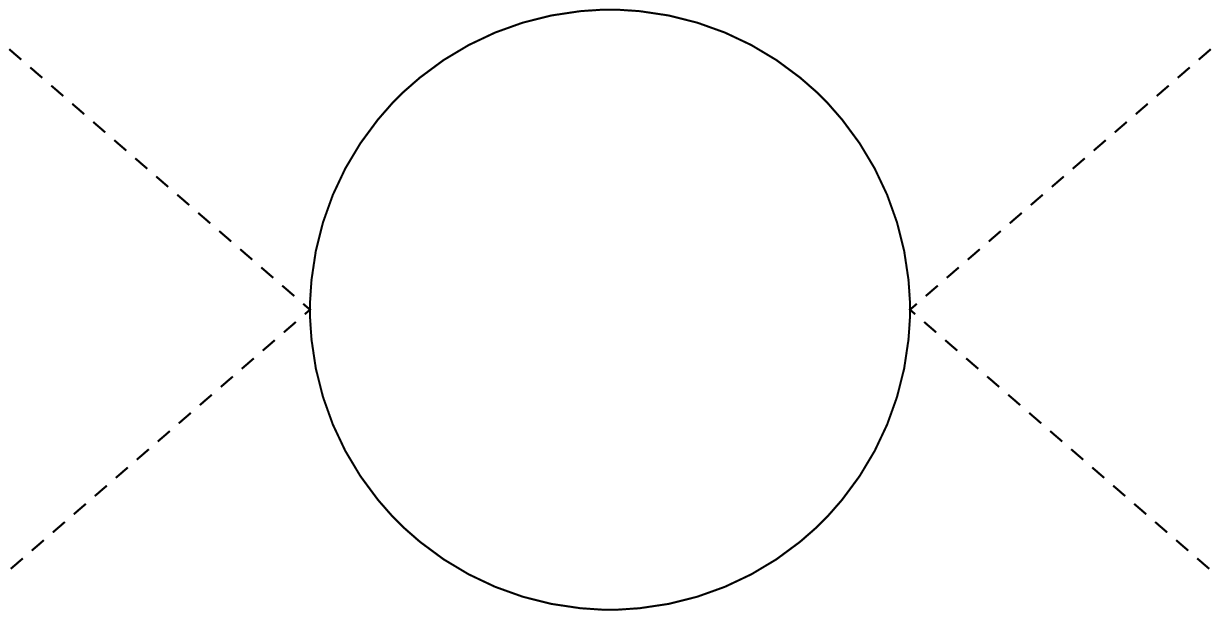}
\hspace{2mm}
\includegraphics[scale=0.25]{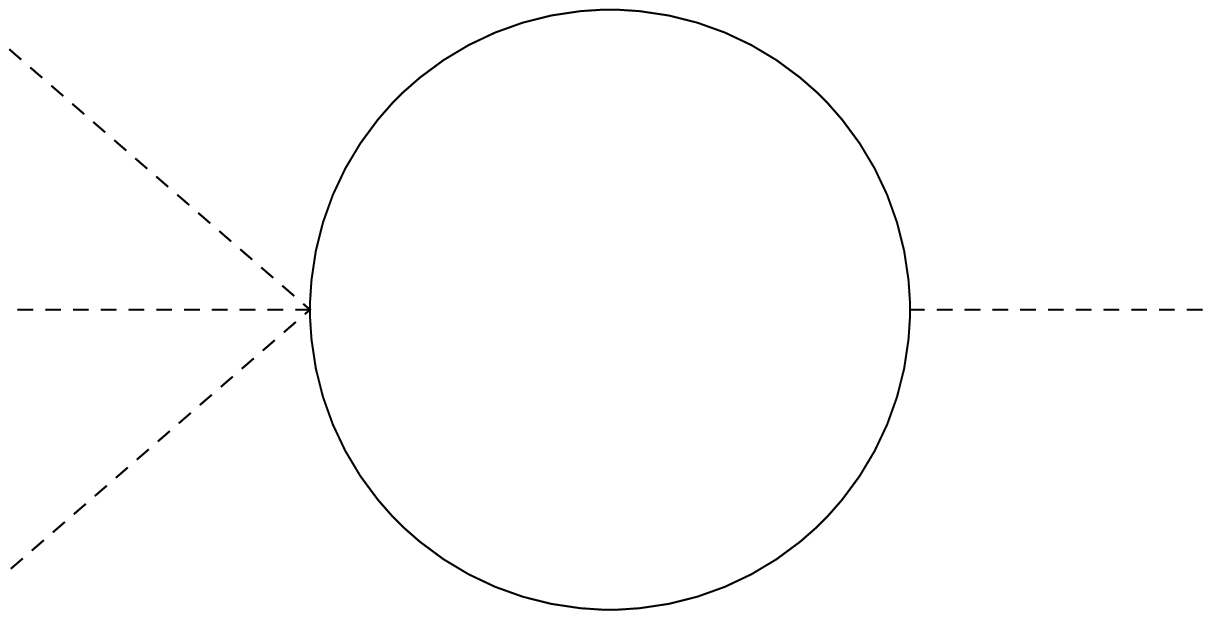}
\hspace{2mm}
\includegraphics[scale=0.25]{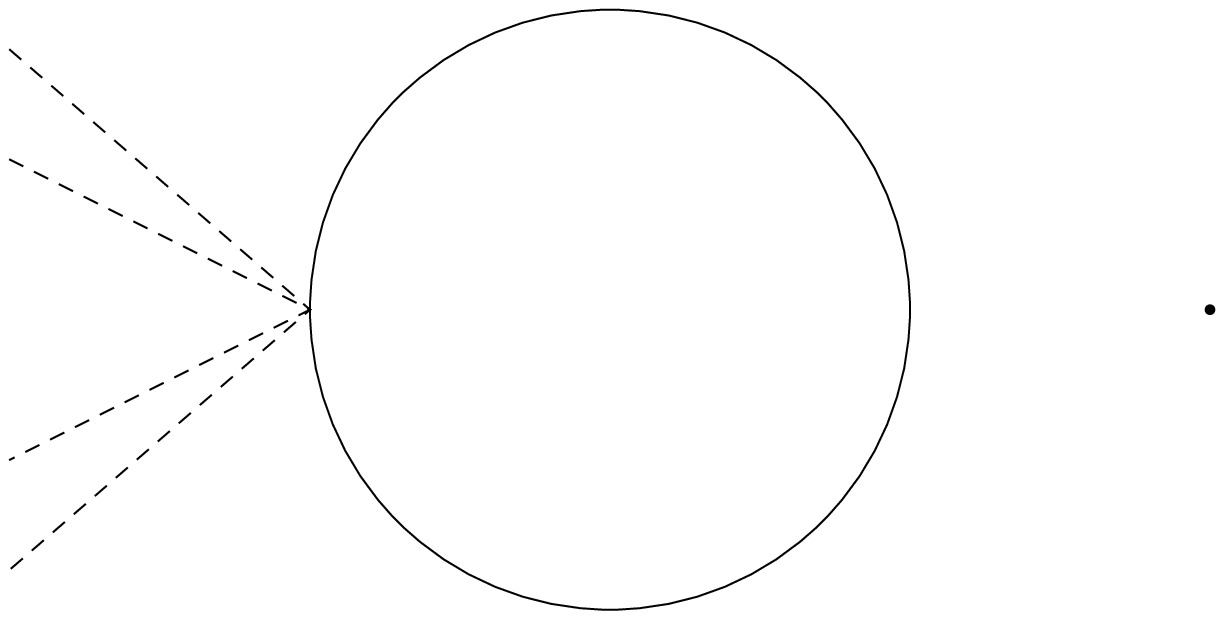}
\caption{Feynman diagram topologies for the four point function of pions
 obtained by integrating out the quarks up to one loop order. Dashed lines
 stand for pions, full lines for quarks.}
\eef{4pt}

Symmetry arguments prevent dimension-3 operators from appearing in
the pion effective action of \eqn{quadpi}. Examination of the Feynman
diagrams which appear to one-loop order in the computation of three-point
functions gives the same result. Technically this happens because we
need to take the trace of Dirac structures such as $\gamma_5$, $\gamma_5
\gamma_\mu \gamma_\nu$ or $\gamma_5 \gamma_\mu \gamma_\nu \gamma_\rho$,
all of which vanish.

The coupling $c^{41}$ in \eqn{quadpi} is computed from the Feynman diagrams
shown in \fgn{4pt}. Since every vertex in these diagrams carries a factor
of $m_0$, the leading term in the chiral limit comes from the last diagram.
This is easily evaluated, and gives
\beq
   c^{41} = \frac{\N m_0}{3f^4} \int\frac{d^4p}{(2\pi)^4}\,\frac m{p^2+m^2}
          = \frac{\N m_0\ppbar}{3f^4} = -\,\frac{m_\pi^2}{3f^2}.
\eeq{c41}
We have used the GMOR relation to get the final form. The evaluation of
the other diagrams is straightforward but tedious.

\goodbreak\section{Describing lattice computations\label{sec:fit}}%\input{sec5.tex}

\subsection{Lattice data set}

We make a few remarks about the lattice computations which we use.
A recent work \cite{brandt} reported two sets of computations of
correlators of the axial current as well as of the pseudoscalar isoscalar
density.  One set (called C1 in \cite{brandt}) uses a quark mass which
gives the crossover temperature $T_{co}=211\pm5$ MeV.  The other set (D1)
has a lighter quark and gives $T_{co}=193\pm7$ MeV. In each scan the
temperature is measured with an accuracy of 2 MeV. However, in the set
D1 the temperature scan below $T_{co}$ mostly covers a range of about
10 MeV of the central value for $T_{co}$, and hence is statistically
indistinguishable from $T_{co}$. As a result, we are forced to use the
set called C1, which covers a larger temperature range. 

Since the EFT is treated in dimensional regularization, loops give
only logarithms of $m/M$ ($M$ is the regularizing scale).  However,
a lattice regularization not only has the corresponding logs of $ma$
(where $a$ is the lattice spacing), but also powers. In the last decade
it was realized that much faster convergence to continuum results are
obtained by subtracting power corrections \cite{bw}. However, 
subtraction of power corrections have not yet been performed for 2-point
functions. Once lattice computations start doing this, a matching with
the EFT will yield continuum results when $T\ll 1/a$.

Comparing our expression for the axial current correlator in \eqn{jjscr}
with that used in \cite{brandt}, we find that what we call $f\sqrt{c^4}$
is called $f_\pi$ there.  Also, what we call the screening mass,
$M_\pi=T_0\sqrt{c^2/c^4}$, is called $m_\pi$ there, and the definition of
the chiral condensate there corresponds to $-\N\langle\bilin{}\rangle$
in our notation. With these translations, the results reported in
\cite{brandt} can be used with the expressions we use. For example, the
GMOR relation is the same, since the factors of $\sqrt{c^4}$ cancel. Also,
the definition of the quantity called $u_f$ in \cite{brandt} is what we
call $u$.

\subsection{Fits}

\bef[bth]
\begin{center}
\includegraphics[scale=0.65]{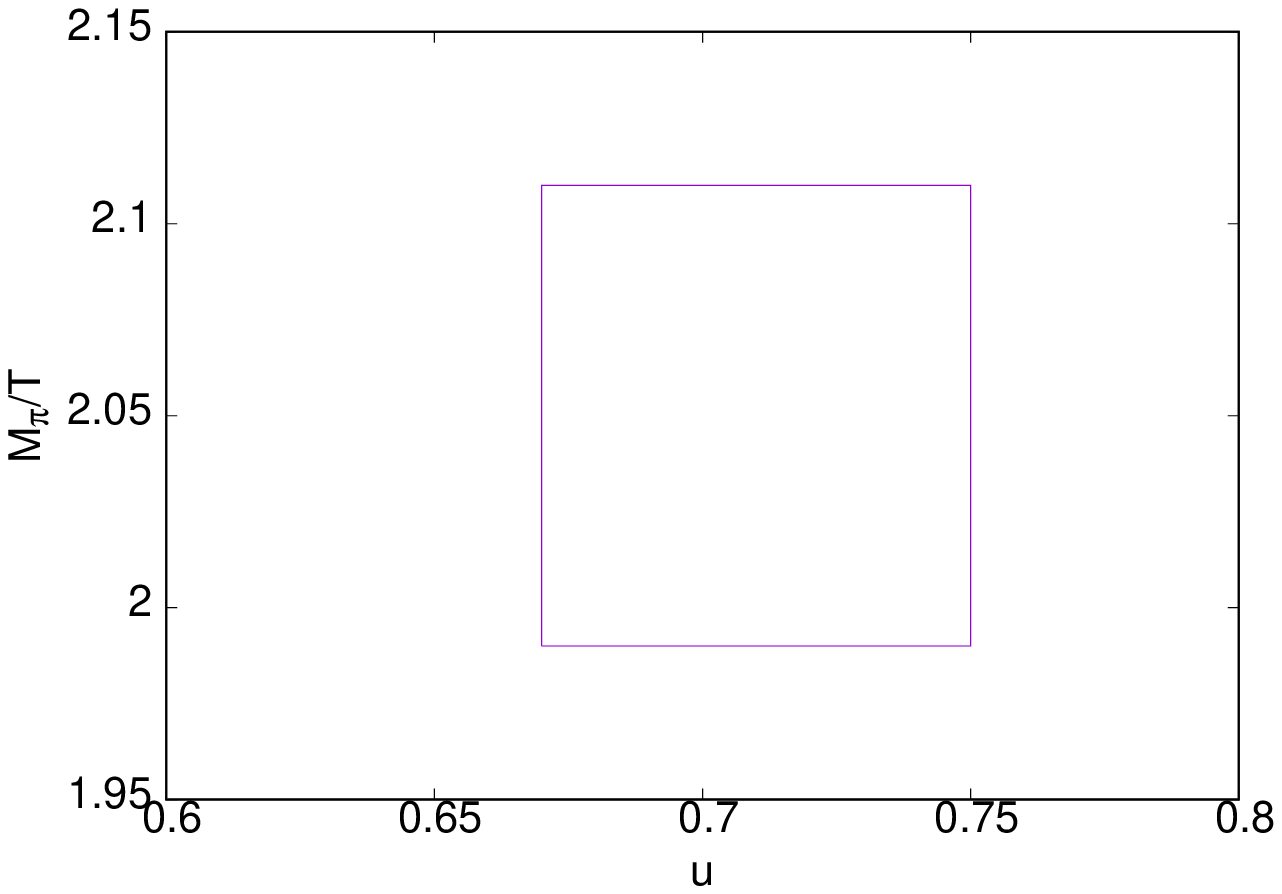}
\includegraphics[scale=0.65]{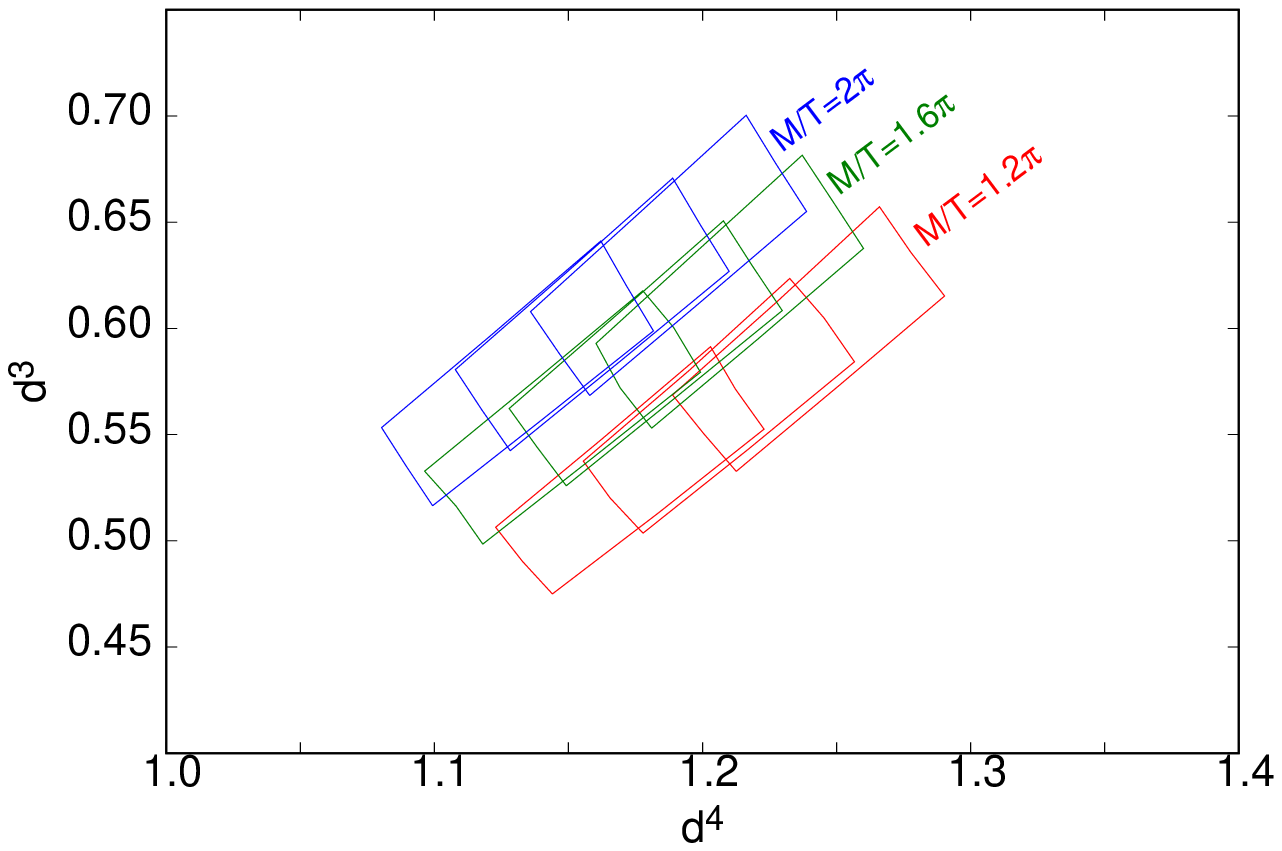}
\end{center}
\caption{The 1-$\sigma$ limits of the lattice results of \cite{brandt}
  at $T/T_{co}=0.84\pm0.02$ define the input rectangular area in the panel
  on the left. The fitted values of the EFT couplings are given in the
  figure on the right. Traversing the rectangle on the left in the clockwise
  direction, one traverses the output rectangles also in the same sense.
  The fits depend on the values of the regularization scale $M$ as shown.
  For each $M$ we have shown three rectangles, these are for the central
  and upper and lower 1-$\sigma$ values of $T_{co}$.}
\eef{extract} 

Since we do not have access to the covariances of the lattice computations
our treatment of the errors is forcibly simplistic. We take the errors
in $M_\pi$ and $c^4$ to be independent, and make four sets of fits at the
extreme values allowed for each of these. Similarly, we take the error in
$T_{co}$ to be completely independent of this.  Also, since we require
the dimensionless ratio $T/T_{co}$, and we have no access to possible
covariances between the numerator and denominator, we add the errors
in quadrature.  If lattice collaborations make these fits in future,
then all the covariances of the inputs can be take into account.

We remarked earlier that there are three parameters of the EFT to be
fitted. Two of these are the dimensionless couplings $d^3$ and $d^4$. The
third coupling is dimensionally transmuted to the value of the critical
temperature $T_c$, which has been chosen to coincide with $T_0$. The
inputs which we use to fix these are the values of $M_\pi$ and $c^4$
(called $m_\pi$ and $u_f$ in \cite{brandt}) at $T/T_{co}=0.84\pm0.02$
(corresponding to $T=177\pm2$ MeV) and the value of the crossover
temperature $T_{co}/T_0$, We define $T_{co}$ in the EFT to be the
temperature at which the chiral susceptibility peaks.

The results of our fits are shown in \fgn{extract}. Little can be said
directly from the values of the couplings, except that they seem to be
of order unity. Since $d^3\to0$ is the chiral limit, this indicates that
the quark mass of the input data set is rather high. We have checked
that the values of $d^4$ obtained in the fit do not violate causality.

The best fit gives $T_{co}/T_0=1.24\pm0.03$. Since the lattice computation
has $T_{co}=211\pm5$ MeV, this yields $T_0=170\pm6$ MeV. Since $T_0$
is also the critical temperature in the chiral limit, this is larger
than expected.  The most likely reason is that all the lattice inputs
are not corrected for powers of $ma$.

\subsection{Checks and predictions}

\bef[htb]
\begin{center}
\includegraphics[scale=0.65]{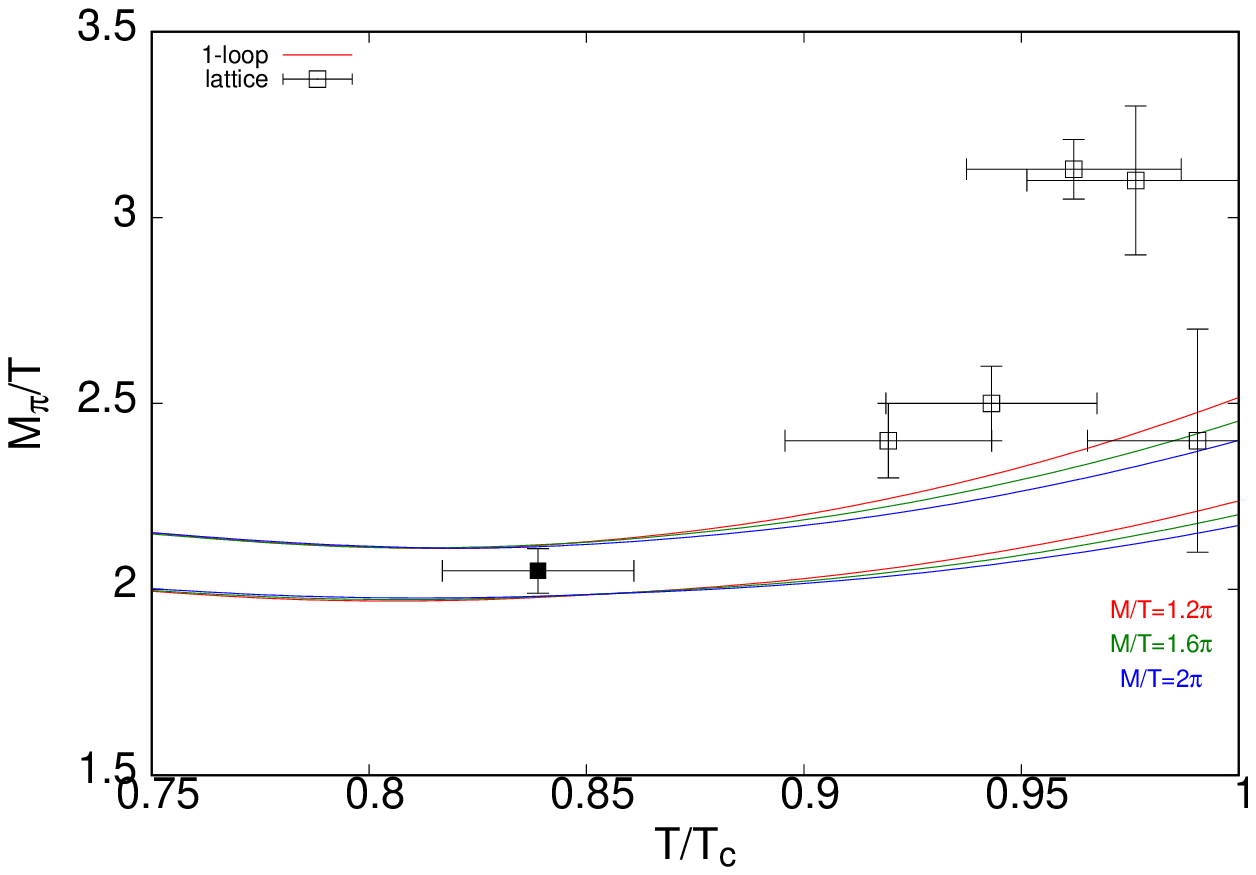}
\includegraphics[scale=0.65]{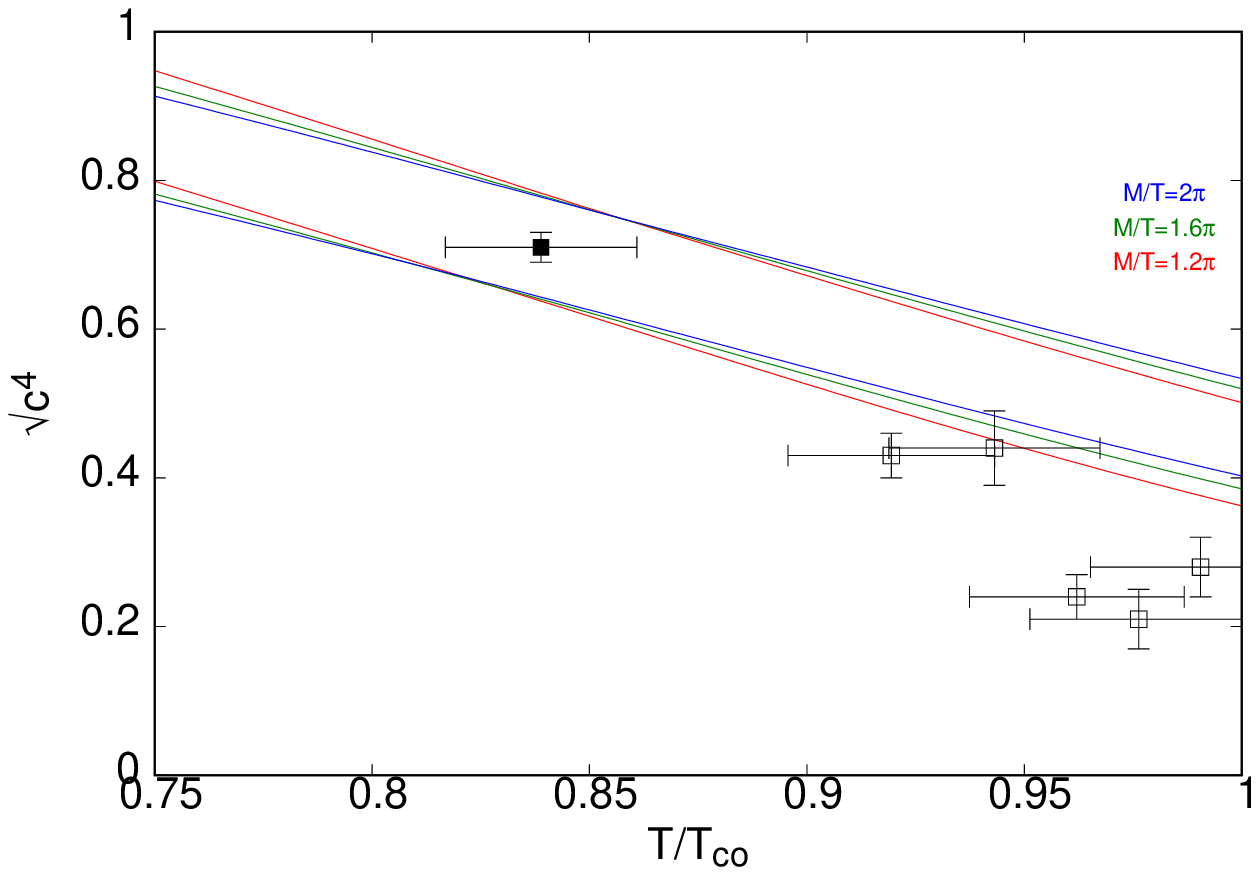}
\end{center}
\caption{Fits to lattice measurements of static pion correlators at finite
  temperature. The lattice measurments which are inputs to the fits are
  shown with filled symbols. Everything else is a check of the EFT. The
  predicted uncertainty bands are shown for different values of $M/T$.
  The legend for $M/T$ on the right edge of the figures shows the ordering
  of the bands, from top to bottom.}
\eef{fits}

\bef[hbt]
\begin{center}
\includegraphics[scale=0.65]{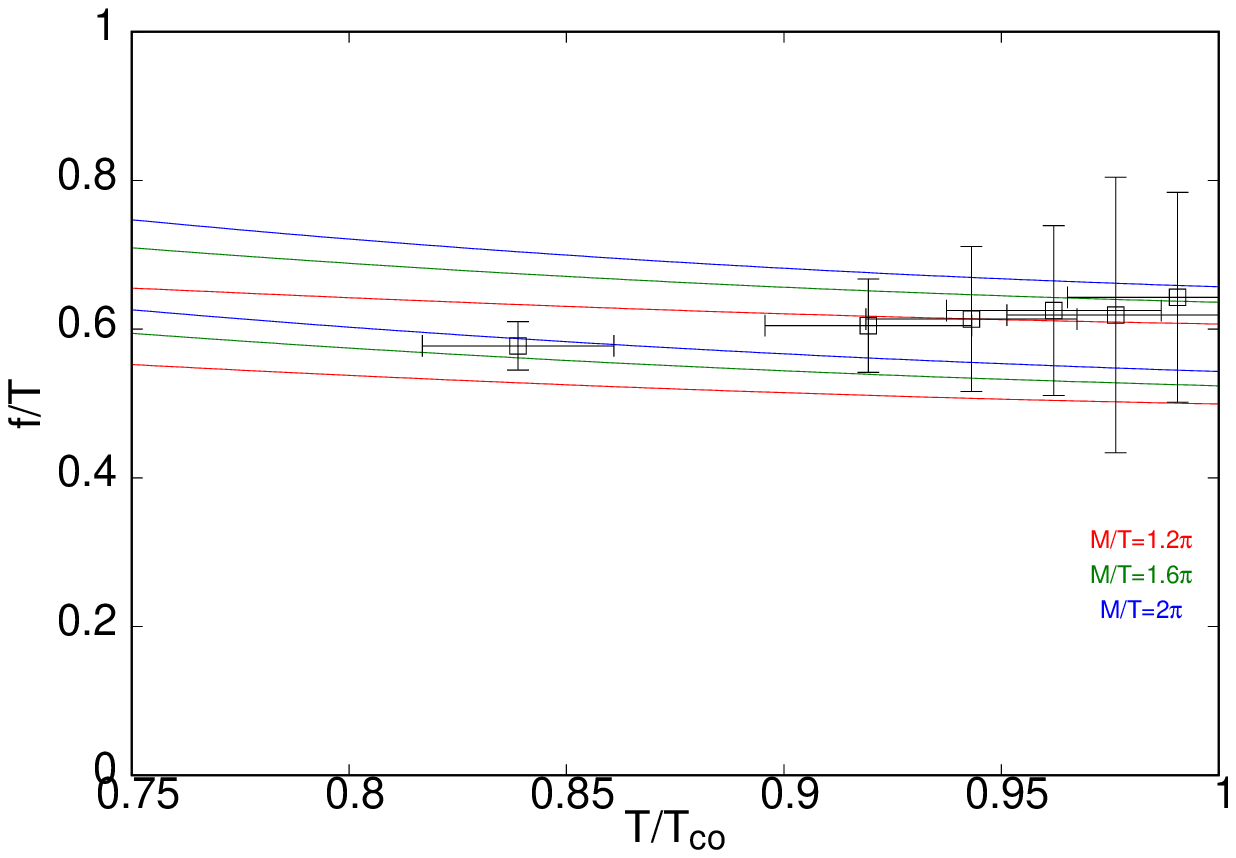}
\includegraphics[scale=0.65]{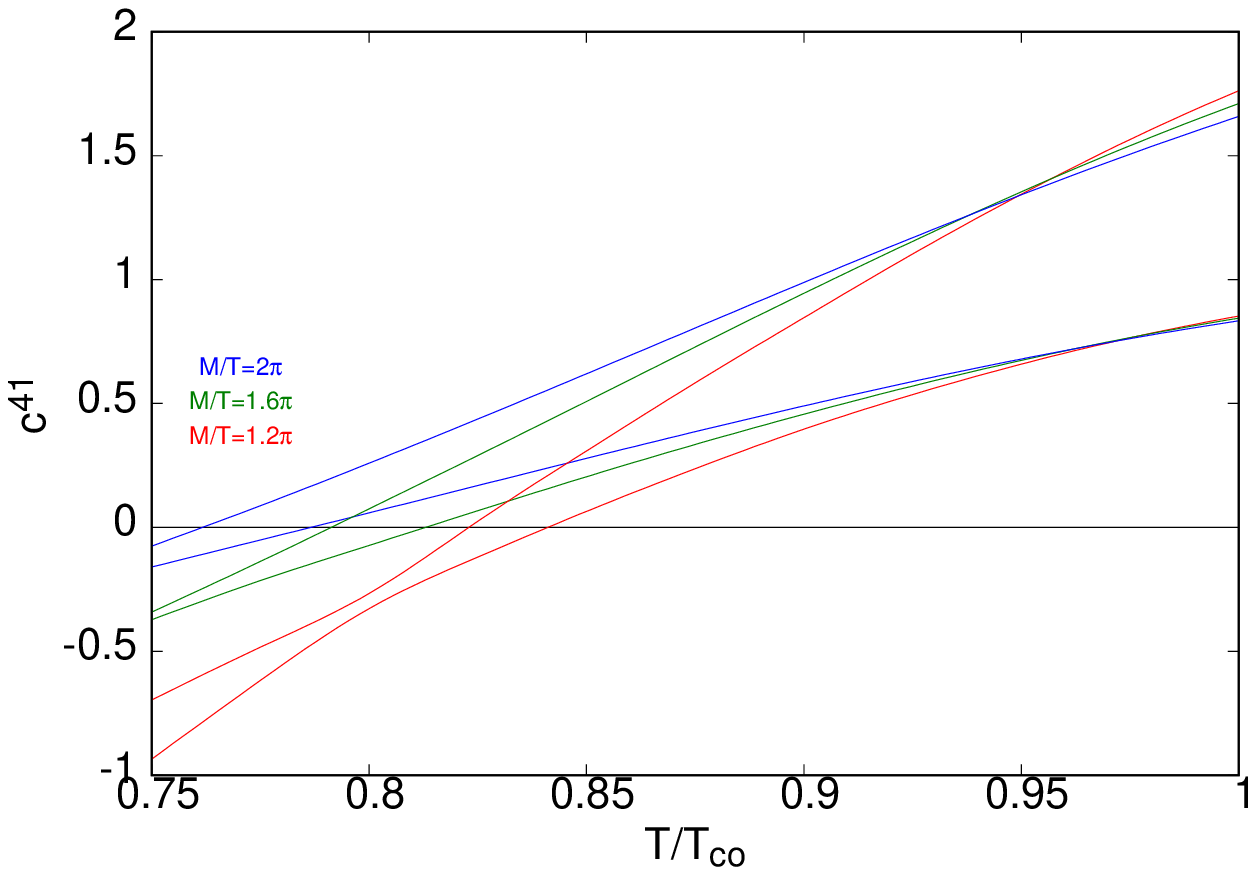}
\end{center}
\caption{From the fitted values of the fermion couplings, other pionic
  couplings can be predicted. The model predictions of $f/T$ and $c^{41}$
  are shown in the two panels. Lattice computations of $f/T$ are compared
  to the EFT predictions.}
\eef{fpi} 

The EFT becomes useful if the couplings $d^3$, $d^4$ and $\lambda$ vary
little with temperature. Then the fit shown in \fgn{extract} can be used
to extract physics at a range of temperatures near $T_{co}$. The first
check of whether this can be done is to examine the temperature dependence
of $M_\pi$ and $c^4$, while keeping the fermionic couplings independent of
$T$. This test is shown in \fgn{fits}. While the agreement is not perfect,
the trend seen in the lattice computation seems to be reasonably well
reproduced in the EFT. The dependence of the EFT predictions on the scale
$M$ is seen to be small. There is some jitter in the lattice computations
which could perhaps be removed if one uses larger statistics. Also,
since the lattice spacing changes as $T$ is changed, at such large quark
masses power corrections could modify the temperature dependance somewhat.

Independent tests of the model are the prediction of the other low-energy
constants: $f$ and $c^{41}$. Since the model parameters are now fixed,
there is no further freedom in these predictions.  A comparison of our
prediction with the lattice results for $f$ is shown in \fgn{fpi}. The
agreement is pleasant. There are no lattice computations of the 4-point
function of pions, so the prediction shown in \fgn{fpi} cannot be tested
now. The dependance of these quantities on the scale $M$ is still mild,
although a little larger than for $M_\pi$ and $c^4$.

\bef[htbp]
\begin{center}
\includegraphics[scale=0.85]{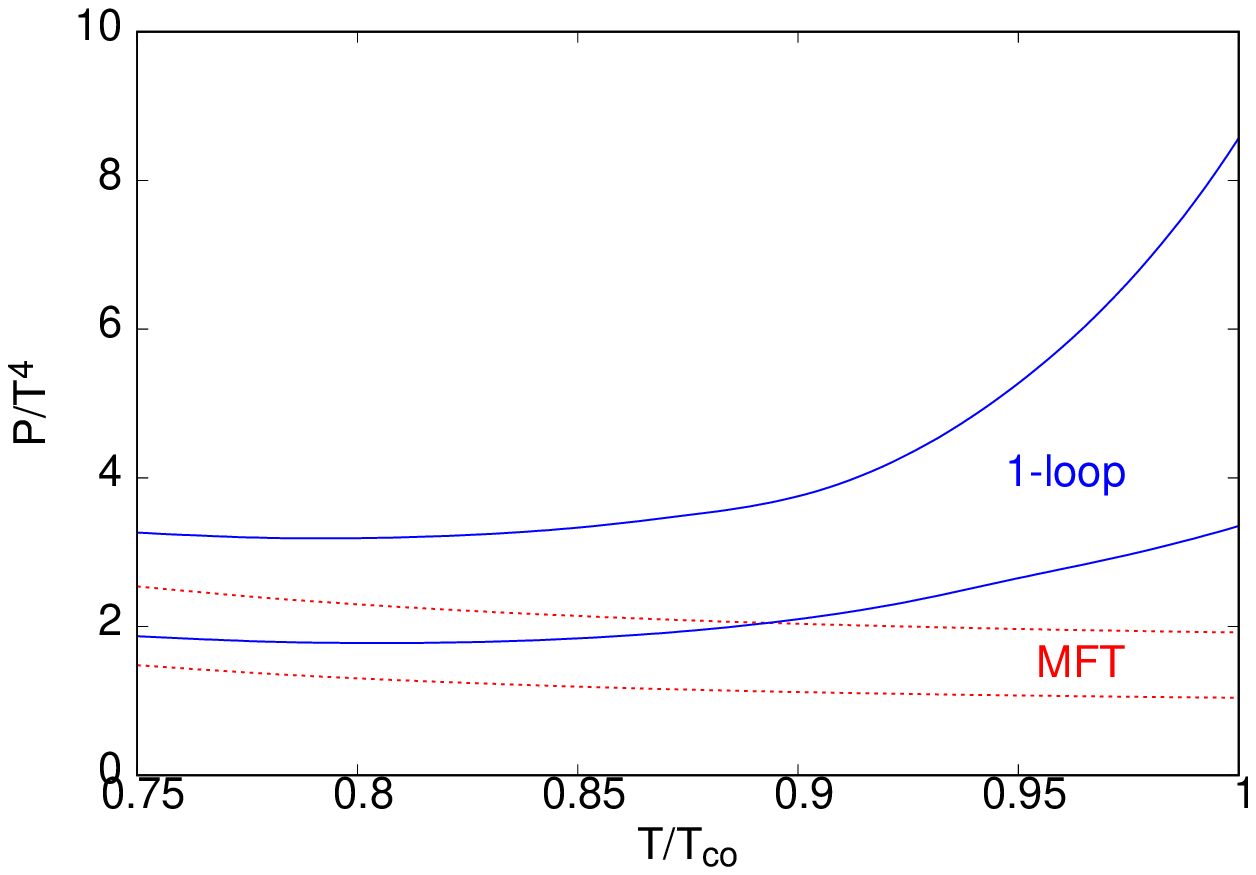}
\end{center}
\caption{From the fitted values of the fermion couplings, thermodynamic
  quantities can be predicted. The model prediction of the pressure,
  $P/T^4$, is shown here.}
\eef{press} 

From these checks it seems that the approximation of neglecting the
temperature dependence of the couplings in the fermion EFT works fairly
well in a range of temperature around $T_c$. The renormalization scale
dependence of physical results is generally smaller than the uncertainty
due to the errors in the input quantities.

However, the results on $c^{41}$ shown in \fgn{fpi} put limits on the use
of the \eqn{quadpi} to compute pion loop corrections in future. Negative
$c^{41}$ means that we are able to treat only fluctuations of magnitude
$\pi<T_0\sqrt{c^2/|c^{41}|}$. In order to do pion loop integrals one
must take into account higher dimension terms which stabilize the path
integral over pions. We also see that at $T\simeq0.75T_c$ the coupling
$c^4$ begins to approach unity.  So it is possible at $T\le0.75T_{co}$, an
effective theory tuned at $T=0$ may become a quantitatively useful tool.

Subject to this limitation, we can now compute extensive thermal
quantities in the effective theory of fluctuations given in
\eqn{quadpi}. This theory is constructed to be valid for $p<T$. For
thermal quantities whose integrands are dominated by momenta in this
region, one should get accurate predictions. Since thermal integrals have
UV cutoffs $\exp(-E_p/T)$, we require $T<\sqrt{c^4T^2+c^2T_0^2}\simeq
T\sqrt{c^4+c^2}$, where we have taken $T_0\simeq T$.  However, for very
small quark masses, near the cross over, where both $c^4$ and $c^2$ are
small, the 1-loop computation of fluctuations cannot give thermodynamic
quantities accurately.  This is connected with the fact that the pion
becomes non-propagating at the critical point in the chiral limit.

In the case we are examining, the convergence criterion is satisfied. Even so,
some quantities are predicted more accurately than others. An examination
of the integrals shows that the energy density, $E/T^4$, or the entropy
density $S/T^3$, are less well controlled than the pressure, $P/T^4$. On
the other hand the chiral susceptibility is better controlled. Recall also
that the pressure is dominated by the lightest particle, and this must
appear in the EFT. Therefore the pressure can be fairly well described as
long as the technical criteria discussed here are satisfied. However the
energy density can get large contributions from massive modes which are
not part of the EFT.

In \fgn{press} we show the pressure computed in the MFT and to 1-loop
order using \eqn{pifene}. The rise seen in the 1-loop correction comes
from a factor of $(c^4)^{-3/2}$ obtained by doing the momentum integrals
in the thermal pion contribution. Since this is a generic feature of the
EFT model, so must be the increase in pressure as one approaches $T_{co}$.

\goodbreak\section{Discussion\label{sec:disc}}%\input{sec6.tex}

In this paper we developed an effective theory for strongly interacting
matter with the modest aim of describing long-distance physics in
a small range of temperatures around the QCD cross over temperature
$T_{co}$. This introduces a scale $T_0\simeq T_{co}$ for the temperature.
Long-distance physics means momenta less than the temperature.

Since QCD at finite quark mass has a cross over, there is no sharp change
in the nature of the degrees of freedom. On one side of the cross over, it
could be natural and easy to use hadronic degrees of freedom, just as it
is natural and easy to use quark and gluon degree of freedom on the other
side. However, the free energy has no singularity, and one should be able
to push either description across the cross over, perhaps with some
increase in the complexity of description.

In this spirit, in \scn{eft} we wrote down the most general Euclidean
effective theory of quarks at finite temperature including dimension 6
terms which are constrained by the chiral SU(2)$\times$SU(2) symmetry of
QCD. Although the action in \eqn{eftaction} is a generalization of the
NJL model, it contains some new features such as the difference between
screening and pole masses. It also includes some extra couplings which
have been considered in the literature from time to time.  An obvious
criticism is that we leave out gluons. We are unable to give a field
theoretical justification for this. Since the model works well enough,
as we showed quantitatively, one should turn this question around and ask
what this implies for possible derivations of effective models from QCD.

The mean field theory is examined in \scn{mft}. The ten couplings for
4-fermi operators reduce to one in this limit, leaving three couplings
to be determined. Divergent integrals are treated in dimensional
regularization (see \apx{reg}). This has a feature which is useful for
thermal physics, namely that it regulates only the integrals which are
divergent, but not the convergent finite temperature pieces.  The gap
equation completely fixes one combination of the couplings, as shown in
\eqn{deftc}, and by dimensional transmutation leaves the transition
temperature in the chiral limit, $T_c$, as the third quantity to be
fixed by data.

There is a critical line in the phase diagram at finite chemical
potential in the chiral limit, whose curvature is predicted in MFT; see
\eqn{curvcrit}. This is within an order of magnitude of the same quantity
computed on the lattice. We discussed in \scn{mft} that a chemical
potential for quarks in the UV theory, namely QCD, gives a hard breaking
of CP symmetry. This means that all operators which break this symmetry
could enter into the EFT. An improved description of the phase diagram
is therefore sufficiently complicated that we leave it to the future.

In \scn{fluct}, we examined low-energy isospin-wave fluctuations around
the MFT. These are organized as another EFT written in terms of pion
fields.  All terms up to dimension-4 are given in \eqn{quadpi}.  Through
the Noether's theorem and PCAC we connected the couplings of the fermion
EFT to lattice computations of two-point functions of currents in QCD.
We demonstrated the 1-loop computation of the couplings of this theory.
We also showed that the Gell-Mann-Oakes-Renner (GMOR) relation remains
valid at 1-loop order in the EFT.

This model is applied to a description of lattice computations in
\scn{fit}.  All the parameters of the model are obtained by matching
the EFT to lattice computations in \cite{brandt} at one value of
the temperature; the results are shown in \fgn{extract}. The lattice
regularization contains power corrections in the quark mass, which
should be subtracted in future. Although the data available today do
not perform these subtractions, the fits yield reasonable descriptions
of the temperature dependence of the long-distance part of two-point
functions of pions at all temperatures given in \cite{brandt}. They also
give predictions for as yet unmeasured four point functions of pions;
see \fgn{fits} and \fgn{fpi}. The pressure shows an interesting rapid
rise below and close to $T_{co}$, which we have argued is a robust
prediction of the model.

The EFT parameters give indications of limits on the model. One limit
comes from the fact that the four-pion coupling $c^{41}$ becomes negative
at about $0.75T_{co}$, indicating that higher dimension terms are needed
to stabilize pion loop corrections. At the same time, $c^4$ becomes of
order unity, indicating that an EFT fitted at $T=0$ may be an appropriate
computational tool at lower temperatures. Near and above $T_{co}$, the
falling value of $c^4$ indicates that higher derivative terms in the
EFT may become necessary. This still leaves a window of applicability
of the 1-loop corrections to the MFT in terms of a pion EFT. The results
we have shown are in this window.

Some future directions are clear. The pion EFT we derived at 1-loop
is very similar to that used in \cite{ss} for examination of departure
from equilibrium.  Similar real-time phenomena may be investigated in
the fermion EFT which we work with. The question of the phase diagram
is another direction, which we have already discussed. A parametrization
of ``radial'' fluctuations can give us a better handle on the physics
above $T_{co}$. The case $N_f=3$ will be interesting. So also will be
a complete analysis of the model at finite chemical potential. One
question that we have not examined at all is of the non-linear sigma
model which describes fluctuations. This can be interesting.

\appendix
\goodbreak\section{Regularization\label{sec:reg}}%\input{app1.tex}

The free energy density for the MFT action in \eqn{emft} can be written as
$\Omega/\N=-T_0^2\Sigma^2/(4\lambda) -I_0(m,T)$, where the integral
\beq
   I_0(m,T) = \int\frac{d^4p}{(2\pi)^4}\,\log(p_4^2+p^2+m^2)
     = T \sum_n\,\int\frac{d^3p}{(2\pi)^3}\,
                 \log\left(\pi^2(2n+1)^2+\frac{E_p^2}{T^2}\right),
\eeq{free}
$E_p^2=(d^4)^2p^2+m^2$, and the sum over Matsubara modes goes
only over positive $n$. This has a cubic divergence which needs to be
cured. The origin of this divergence will be important to understand,
since we expect no divergences in the thermodynamics of free fermions.

Using a standard trick \cite{kapusta} we can write
\beqa
\nonumber
   I_0(m,T) &=& T \int\,\frac{d^3p}{(2\pi)^3}\,\sum_n\,\left[
                 \log\left(1+\pi^2(2n+1)^2\right) +
      \int_1^{E_p^2/T^2} \frac{dy}{\pi^2(2n+1)^2+y}\right] \\
        &=& T \int\,\frac{d^3p}{(2\pi)^3}\,\left[\sigma_l+
      \frac1{\pi^2}\int_1^{E_p^2/T^2} dy\sum_n\frac1{(2n+1)^2+y/\pi^2}\right],
\eeqa{simple}
where we have introduced the notation $\sigma_l$ for the temperature
independent but divergent sum over the logarithmic terms. We may justify the
interchange of the sum and the integral by putting an arbitrary UV cutoff at
all stages until the cubic divergence in terms of this cutoff is removed and
a finite result is obtained. After the change of variables, $t=\sqrt y$, and
performing the remaining sum, one gets \cite{kapusta}
%\beq
%   I_0(m,T) = T \int\,\frac{d^3p}{(2\pi)^3}\,\left[\sigma_l+
%             \frac12\int_1^{E_p/T}\,dt\tanh\frac t2\right],
%\eeq{simpler}
%As a result, we have
\beq
   I_0(m,T) = T \int\,\frac{d^3p}{(2\pi)^3}\,\left[\sigma_l'+
               \frac{E_p}{2T} + \log\left(1+{\rm e}^{-E_p/T}\right)\right],
\eeq{atlast}
where $\sigma_l'$ absorbs all the factors independent of $p$ and $T$. Note
that there are now two kinds of cubic divergences: one is linear in $T$
and comes from the integral over $\sigma_l'$, the other is independent of
$T$ and comes from the integral over $E_p$. The second term is just the
zero-point energy of the fields, called the vacuum energy.  The first
term, which we may call the vacuum entropy, gives no contribution to
derivatives of $\log Z=-F/T$, and hence cannot give any contribution to
thermodynamics, and so may be subtracted out.

A cutoff regularization is easy to implement
when it works, and cures all UV divergences. However, while it makes the
vacuum entropy finite, it does not reduce it to zero. In view of this, a
different regularization may be more useful. An alternative is to use
a scheme which resembles Pauli-Villars regularization.
In the familiar version of this regularization process one subtracts
from a divergent integral another integral of the same form with $m\to
M$ and choosing $M>>m$. However, in this case that would still leave a
linear divergence. We are forced to subtract more integrals to obtain
a regular integral. This removes the vacuum entropy term. However the
vacuum energy has powers of $m/M$.

In zero-temperature effective theories, dimensional regularization turns
out to be very useful. At finite temperature one would have to work
near $D=3$ spatial dimensions, since we want to regulate after
doing the sum over all Matsubara frequencies. Recall that dimensional
regularization needs a regularization scale $M$ which is used to
give the correct dimension to the integral defined originally with
$\epsilon=0$. The basic formula we need to use is
\beq
   J^m_n = \mu^{3-D}\int\frac{d^Dk}{(2\pi)^D}\,\frac{k^{2m}}{(k^2+\ell^2)^n}
    = \ell^{2m+D-2n}\mu^{3-D}\,\frac{\Omega_D}{(2\pi)^D}
        \,\frac{\Gamma(m+D/2)\Gamma(n-m-D/2)}{2\Gamma(n)},
\eeq
where $\Omega_D=2(2\pi)^{D/2}/\Gamma(D/2)$ is the volume of an unit
sphere in $D$ dimensions. We set $D=3-2\epsilon$.

For the vacuum entropy term, we have $m=0$ and $n=0$. The formula
above shows that the integral is finite, and therefore zero due to its
invariance under the choice of $\ell$.  The vacuum energy term requires
$m=0$ and $n=-1/2$. The formula then gives the following result
\beq
   J^0_{-1/2} = \left(\frac{\ell^4}{32\pi^2}\right)\left[
    -\frac1\epsilon+\gamma-\frac32+\log\left(\frac{\ell^2}{4\pi\mu^2}\right)
    +{\cal O}(\epsilon)\right],
\eeq{eneepsilon}
where $\gamma$ is the Euler-Mascheroni constant.

In order to compute the vacuum energy we scale $p\to(d^4)p$ in
\eqn{atlast}.  However, the regularization scale $\mu$ is a scale
on $p$, so it must be scaled in the same way.  Using the definition
$M^2=4\pi\mu^2\exp(-\gamma)$, and dropping the pole in $\epsilon$,
we then get the $\overline{\rm MS}$ result
\beq
   I_0^r(m) = -\left(\frac{m^4}{64\pi^2(d^4)^3}\right)
    \left[\log\left(\frac{m^2}{(d^4)^2M^2}\right)-\frac32\right].
\eeq{drvene}

Putting everything together, we find
\beq
   I_0(m,T) = I_0^r(m) + \frac{T}{(d^4)^3} \int\,\frac{d^3p}{(2\pi)^3}
     \left[\log\left(1+{\rm e}^{-\sqrt{p^2+m^2}/T}\right)\right]
\eeq{regularagain}
One can continue the thermal integral to arbitrary dimension.  Since the
integral is finite, taking the result to the limit $D\to3$ gives no
poles in $3-D$, and hence does not introduce the scale $M$ into the
thermal integral. DR is therefore much more intuitive for the thermal
part than either a cutoff or Pauli-Villars regularization would have
been. One pleasant result of this is that large values of $\log(M/m)$
can be easily avoided, in the same way as at zero temperature. DR has been
used before at finite temperature in \cite{inagaki}.

Using \eqn{pmft}, the gap equation, $d\Omega/d\Sigma=0$, becomes
\beq
   \Sigma= -\,\frac{2\lambda}{T_0^2}\left\{\frac{m^3}{16(d^4)^3\pi^2}
         \left[\log\left( \frac{m^2}{(d^4)^2M^2}\right)-1\right]
       +\frac m{2(d^4)^3\pi^2}\int_0^\infty\frac{dp p^2}E\,\frac1{\exp(E/T)+1}
         \right\}.
\eeq{dmft}
In the chiral limit, $d^3=0$ and so $m=\Sigma$. We can see then
that $\Sigma=0$ is always a solution, and that there are generally two
more real solutions related by sign flips.

Pion loop integrals may also be regularized in DR. We see that
the convergent thermal parts of the integrals will give results of the
form $T^nf(c^2 T_0/T)$, where $n$ is the engineering dimension of the
loop integral. At finite temperature this introduces a new scale in
amplitudes, which would then modify the power counting of pion loops
and give results different from those in, for example, \cite{wein}.
This will be interesting at higher loop orders than what we examine.

\end{document}